\documentclass[pra,reprint,groupedaddress,floatfix,superscriptaddress]{revtex4-2}

\usepackage{graphicx}%
\usepackage{braket}
\usepackage{amsmath,amssymb,amsfonts}%
\usepackage{multirow}%

\usepackage[pdfencoding=auto, psdextra]{hyperref}
\hypersetup{colorlinks=true,urlcolor=blue, citecolor=blue, linkcolor=blue}

\begin{document}

\title{Measuring the evolution of entanglement in Compton scattering}

\author{Igor Tkachev}
\email{tkachev@inr.ru}
\affiliation{Institute for Nuclear Research RAS, Moscow, 117312, Russia}
\author{Sultan Musin}
\affiliation{Institute for Nuclear Research RAS, Moscow, 117312, Russia}
\affiliation{Moscow Institute of Physics and Technology, Moscow, 141701, Russia}
\author{Dzhonrid Abdurashitov}
\affiliation{Institute for Nuclear Research RAS, Moscow, 117312, Russia}
\author{Alexander Baranov}
\affiliation{Institute for Nuclear Research RAS, Moscow, 117312, Russia}
\affiliation{National Research Nuclear University MEPhI, Moscow, 115409, Russia}
\author{Fedor Guber}
\affiliation{Institute for Nuclear Research RAS, Moscow, 117312, Russia}
\author{Alexander Ivashkin}
\affiliation{Institute for Nuclear Research RAS, Moscow, 117312, Russia}
\author{Alexander Strizhak}
\affiliation{Institute for Nuclear Research RAS, Moscow, 117312, Russia}
\affiliation{Moscow Institute of Physics and Technology, Moscow, 141701, Russia}

\date{\today}

\begin{abstract}
The evolution of the entanglement measure during Compton scattering is studied. Our analytical results show that the corresponding measure coincides with the concurrence of the two-qubit state arising after scattering. The state never collapses to a separable one, contrary to what was  previously assumed. The behavior of quantum entanglement during scattering is identical to the behavior of initially classically correlated photons up to a constant factor equal to two. This is consistent with local quantum field theory, and "spooky action at a distance" is not required to explain the change in state of nonlocally entangled qubits during the measurement of one of them. Our dedicated experiment with annihilation photons confirms these results and explains the "Puzzle of Decoherence'' observed recently.
\end{abstract}


\maketitle

\section{Introduction}

The behavior of quantum correlations in entangled states during measurements is orthogonal to the classical perception of reality and is crucial for testing the foundations of quantum mechanics. The inherent nonlocality of such states, first noted by Einstein, Podolsky, and Rosen (EPR)~\cite{EPR-paradox}, has led to paradoxes, the concepts of spooky action at a distance (SAD) and quantum steering~\cite{Schroedinger1935}, and hidden variable theories as an alternative to quantum mechanics.  The latter can be tested~\cite{Bell, Bell2} and was disproved in favour of quantum mechanics~\cite{Aspect2}. By steering, Schrödinger meant the ability of an observer to influence the state of a distant system through local measurements. In its modern formulation, EPR-steering has been experimentally demonstrated for Bell-local entangled states~\cite{Saunders2010}, while recent  experiments have placed limits on the speed of SAD during photon pair entanglement measurements that exceed the speed of light by at least four orders of magnitude~\cite{Salart2008,SantamariaAmato2023}. 
 
Historically,  the very first entangled state was obtained as a pair of photons produced by the positron-electron annihilation at rest~\cite{Wheeler}. According to the quantum theory~\cite{Pryce,Snyder}, two annihilation photons are in maximally entangled state with mutually orthogonal polarizations. 
Back in the middle of the last century, Bohm and Aaronov~\cite{Bohm} realized that these photons make it possible to  approach the  EPR problem experimentally. Their pioneering idea to measure the enhancement in azimuthal angular correlation of Compton scattered annihilation photons was implemented in a series of experiments~\cite{Langhoff, Faraci, Kasday, Bertolini1981, Wilson1976, Bruno77} that proved the entangled nature of annihilation photons. Following the discovery of spontaneous parametric down-conversion, the focus of entangled state research shifted to quantum optics and formed the basis of quantum technologies such as quantum communications and computing. 

Interest in annihilation photons was revived  by Bohm and Hiley~\cite{Bohm2}, where they argued that after measuring one of the photons, a pair of annihilation photons collapses into a mixed separable state. The authors concluded that these decoherent photons completely lose their azimuthal angular correlation and hence can be easily distinguished from the initial ones. 
This is potentially important for  some modern applications. In particular, quantum entangled positron emission tomography (QE-PET)  is currently being actively developed by the international scientific community~\cite{watts,McNamara,Toghyani,kozuljevich}.  For other possible technological applications of entangled penetrating radiation, ranging from medical to   industrial imaging, see  review~\cite{Ujeniuc2024}. 

In recent work~\cite{Abdurashitov,Ivashkin}, it was discovered, and confirmed in~\cite{Parashari2023}, that correlations of annihilation photons, surprisingly, do not disappear after one of them has been scattered (measured). This observation was called the "Decoherence Puzzle'' in~\cite{J-PET2,Parashari2023}. Collapse into a  separable state does not occur.  
The discovery could have important implications for the foundations of quantum mechanics. And in any case, it is important for technological applications to understand how the inevitable scattering processes affect entangled photons of any energy~\cite{Cande2014,Valencia2020,Safadi2023}.  

In this paper we study the behavior of entanglement during polarization measurements (aka Compton scattering in the case of annihilation photons). Our approach is threefold: analytical, Monte-Carlo simulations of classically correlated photons, and real experiment.  The results of the first two approaches are valid at any energy, and the actual experiment was carried out for annihilation photons.

\section{Correlations, their visibility and detection efficiency}

Entanglement reveals  itself as directly observable interferometric visibility.  However, at high energies, unlike, say, optics, visibility is degraded due to the polarimeter's intrinsic efficiency, which varies from zero to one depending on the scattering angle in the Compton polarimeter, so additional effort is required to extract entanglement from visibility. Fortunately,  this is possible, as shown below.

\subsection{Efficiency of a Compton polarimeter}

The study of polarization states of annihilation photons can be performed only by Compton polarimeters. The operation of such polarimeters is based on the dependence of the differential cross section  (probability P) of Compton scattering on the polarization direction, which is given by the Klein-Nishina formula~\cite{Nishina}:
\begin{equation}
d\sigma/d\Omega = P = r_e^2/2\cdot(E_f/E_i)^2\cdot(\gamma - 2\sin^2{\theta}\cdot\cos^2{\phi}),
\label{eq:kl-ni}
\end{equation}
where ${r_e}$ is the classical electron radius, $E_i$ is the energy of the incident photon, $E_f$ is the energy of the scattered photon, 
\begin{equation}
\gamma \equiv \frac{E_f}{E_i}+\frac{E_i}{E_f}, \qquad \cos{\theta} = 1 + \frac{m_e}{E_i} - \frac{m_e}{E_f}, 
\label{eq:sangle}
\end{equation}
$\theta$ is the angle between the incident and scattered photons momenta forming the scattering plane,  $\phi$ is the angle between the scattering plane and the direction of polarization of the incident photon, and, finally, $m_e$ is an electron mass. As follows from the equation~(\ref{eq:kl-ni}), photons are scattered predominantly orthogonally to the plane of polarization. This property is used to measure the polarization of high energy photons using Compton polarimeters.

The efficiency of a Compton polarimeter is expressed through its analyzing  power as $A \equiv \frac{N_\perp-N_\parallel}{N_\perp+N_\parallel}$, where ${N_\perp}$ $({N_\parallel})$ denotes the number of detected events in counters located perpendicular (parallel) to the polarization of initial photons. Using Eq.~(\ref{eq:kl-ni}) the analyzing power might be analytically expressed as~\cite{Knights}:
\begin{equation}
A =\frac{P(\theta,\phi=90^\circ)-P(\theta,\phi=0^\circ)}{P(\theta,\phi=90^\circ)+P(\theta,\phi=0^\circ)} 
=\frac {\sin^2{\theta}}{\gamma -\sin^2{\theta}}
\label{eq:apow}
\end{equation}

As follows from the definition and Eq.~(\ref{eq:sangle}), the analyzing power can be considered as a function of $E_i$ and $\theta$, or $E_i$ and $E_f$. For a given energy of the initial photon, the efficiency of a polarimeter depends on the scattering angle and for $E_i=511$~keV reaches maxim of $A = 0.69$ at $\theta=82^\circ$ and at $\theta=90^\circ$  is a few percent lower,  $A = 0.67$ . 

At least two Compton polarimeters are required for measuring angular correlations of annihilation photon pairs. We refer to the particles in the pair as photons ``a" and ``b". Each elementary polarimeter consists of a Compton scatterer and two detectors of scattered gammas arranged orthogonally. Like a conventional visible light polarimeter, the scatterer measures the polarization of high-energy annihilation photons, and scattered photon detectors act as photodetectors in optical polarimeters.

\begin{figure*}
\centering 
\includegraphics[width=0.5\textwidth]{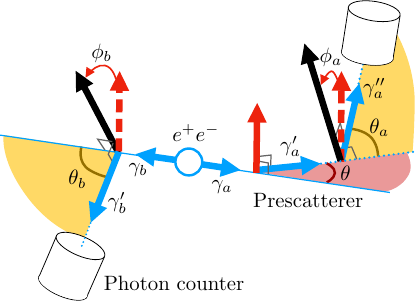} \hspace{0.5cm} 
\includegraphics[width=0.25\textwidth]{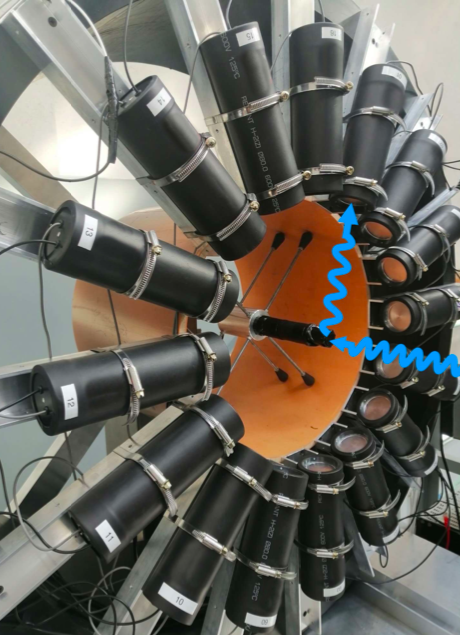}
\caption{\label{fig:i}  Left:  Schematic diagram of a two-arm experimental setup for measuring polarization correlations of annihilation photons. Each arm consists of a pre-scatterer and a main scatterer, followed by photon counters.  The  source of positrons is placed in a lead collimator between the arms. Blue arrows indicate photon momenta, black arrows indicate normals to the scattering planes in the polarimeter, and red arrows indicate normals to the pre-scattering plane. For forward scattering $\phi = \phi_a+\phi_b$ is the azimuthal angle between the signal photon counters. 
Right: Photo of the experimental setup arm. The blue line shows the path of the photon scattered in the main scatterer and absorbed in the counter.
}
\label{scheme_polarimeter}
\end{figure*}

In what follows we study the correlations in the state when one of the photons, either ``a" or ``b", was scattered before entering the polarimeter.  For clarity in what follows, let this happen for photon "a". For a scattering event occurring in a particular polarimeter, we use notations $E_{ai}$, $E_{af}$, $\theta_a$ and  $E_{bi}$, $E_{bf}$, $\theta_b$,  while for the prescattering event $E_{i}$, $E_{f}$, $\theta$ are used, see Fig.~\ref{scheme_polarimeter}. Clearly $E_{ai} = E_{f}$ and for annihilation photons $E_{bi}= m_e$.


\subsection{Visibility of correlations}

When polarizations of two photons are correlated, their counts in two directions separated by an angle $\phi$  will be modulated in general  as
\begin{equation}
N(\phi) = P_0 \left (1 -  \nu\, \cos2\phi \right),
 \label{eq:vis_def}
\end{equation}
for the particular case of  annihilation photons see e.g.~\cite{Pryce,Snyder}.
The parameter $P_0$ is irrelevant in what follows, but $\nu$, which is called {\it visibility}, contains  both the correlations in the initial two-photon state and  the efficiencies of  the polarimeters. The simplest way to extract $\nu$ from data using a small number of polarimeters is to arrange the counters in multiples of $\phi=90^\circ$ and calculate $\nu = \left( N(90^\circ) - N(0^\circ)\right) /(N(0^\circ) + N(90^\circ)) $. In our experimental setup we are utilising 16 photon counters covering $360^\circ$ circle of each arm, see Fig.~\ref{scheme_polarimeter}. Therefore, we directly fit Eq.~(\ref{eq:vis_def}) to data to extract $\nu$. Alternatively, in our setup we can fit counts to the S-function of Clauser-Horne-Shimony-Holt (CHSH) inequality~\cite{CHSH}
\begin{equation}
 S = \nu\, (\cos6\phi - 3\cos2\phi )
 \label{eq:Sfunction}
\end{equation}
producing the same $\nu$ but with better accuracy.

\subsection{Untangling entanglement from visibility}

The positron and electron annihilate predominantly with zero relative angular momentum into a two-photon state. The resulting two photons have opposite momenta and an energy of 511 keV, equivalent to the mass of an electron. Due to the parity and angular momentum conservation the two-photon polarization state is described as a combination of left and right hand circularly polarized photons 
$ {\Phi^-} = ( \ket{{L_a}{L_b}} - \ket{{R_a}{R_b}})/{\sqrt2}$.
Since Compton scattering is sensitive to linear polarization, it is convenient to rewrite this state as
\begin{equation}
 {\Psi^+} = \frac{1}{\sqrt2}( \ket{{H_a}{V_b}} + \ket{{V_a}{H_b}}),
 \label{eq:wavefunc2}
\end{equation}
where H(V) represents the horizontal (vertical) linear polarization of the first or second photon. $\Psi^+$ is one of Bell's four maximally entangled basis states. The visibility of the azimuthal angular correlations between these two Compton scattered photons~\cite{Pryce,Snyder} can be rearranged as  
$$\nu (\Psi^+)=A(E_{ai},\theta_{a})\, A(E_{bi},\theta_{b}), $$
see e.g.~\cite{Caradonna,Hiesmayr}.
$\Psi^+$ can be compared to the state of classically correlated photons. By construction, photons in the latter pairs initially have mutually perpendicular polarizations in randomized directions. For this classically correlated  state, using Eq.~(\ref{eq:kl-ni}), it is easy to find that 
$$\nu_{cc}=\frac{1}{2} A(E_{ai},\theta_{a})\, A(E_{bi},\theta_{b}), $$ for derivation see section~\ref{sec:CC}. 

This factorization of visibility into the efficiency of polarimeters should not be specific to the states considered. Indeed, it can be expected that the appropriate measure of entanglement (characteristic of the state being measured) and the efficiency of the polarimeter (characteristic of the measuring device) should factorise  for these notions to make sense.  Based on this expectation, for the general two-photon state $\rho$ measured in a polarimeter, we write
\begin{equation}
 \nu (\rho) = C(\rho)\, A(E_{ai},\theta_{a})\, A(E_{bi},\theta_{b}) ,
 \label{eq:MainRel}
\end{equation}
where $\rho$ is the polarization density matrix of this state, $C = 1$ for a maximally entangled state and cannot exceed 1,  as we saw above, and for a separable state $C = 0$.  This relation will be proven below for the case of 3-Compton scattering with $C$ being a concurrence, a particular measure of entanglement~\cite{PhysRevLett.80.2245}, cf.~\cite{PhysRevA.104.013704, PhysRevLett.130.090202} for the case of optical photons. 

Using results of a recent paper~\cite{Caradonna2024}, which is based on  Quantum Field Theory (QFT) in matrix representation for both the photon and the interacting Compton electron~\cite{Fano1949,McMaster1961},    we found that the visibility of correlations of the initially maximally entangled pair after three Compton scattering process do factorises  as 
$$\nu = C(E_i,\theta)\, A(E_{ai},\theta_{a})\, A(E_{bi},\theta_{b}), $$ 
see Sec.~\ref{sec:Fact}. 
This factorisation  allows us to measure the change of entanglement in the scattering event and infer the polarization state of a photon pair after scattering.

As a measure of entanglement, we  found, see Sec.~\ref{sec:Fact}
\begin{equation}
\begin{aligned}
 &C( \theta, E_i) =   \frac{ (1+\left| \cos\theta \right|)^2} { 2 (\gamma -\sin^2{\theta})}, \\
 &C( \theta, m_e)= \frac{1+\rm{abs}(\cos\theta) -  \frac{1}{2} \sin^2 \theta}{1+\cos^2 \theta + (1-\cos \theta )^2/(2-\cos \theta )} \, ,
 \label{eq:CC}
\end{aligned}
\end{equation}
where in the second equality we used $E_i = m_e$, displaying the explicit angular dependence for this case. The first expression is true at all energies, but at low energies $\gamma = 2$, which  gives an explicit angular dependence of the entanglement measure at $E_i \ll m_e$.

\begin{figure*}[hbtp]%
  \centering
 \includegraphics[width=0.495\textwidth]{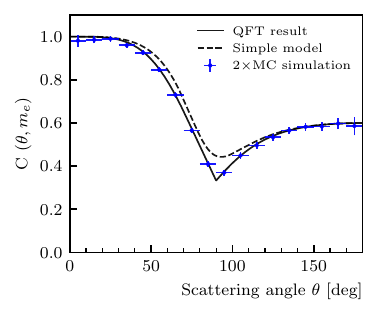}
 \includegraphics[width=0.495\textwidth]{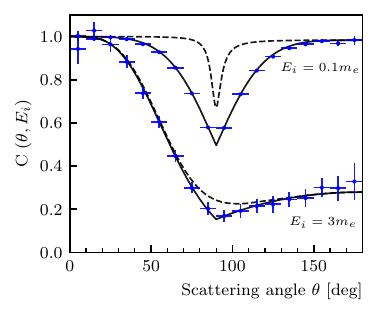}
  \caption{
A measure of entanglement predicted by QFT, Eq.~(\ref{eq:CC}), is shown with solid lines for three values of  initial photon energy, $E_i = m_e$ (left) and $E_i = 0.1\, m_e$ and  $E_i = 3\, m_e$ (right). One of the photons of the initially maximally entangled pair experienced Compton scattering at an angle $\theta$. The dashed lines represent a simple model with the resulting concurrence given by Eq.~(\ref{eq:C2}). The blue dots with error bars represent twice  the results of  Monte Carlo simulation of initially classically correlated photons, where $C(E_i,\theta)$ was extracted using Eq.~(\ref{eq:MainRel}). Here and in Fig.~\ref{fig:ConcrData}, the horizontal error bars indicate bin sizes in $\theta$. 
}
\label{fig:ConcrMC}
\end{figure*}

The function Eq.~(\ref{eq:CC}) is shown in Fig.~\ref{fig:ConcrMC} by a solid line for three values of $E_i$. Note that $C( \theta, E_i)$ for $E_i = 0.1\,m_e$ already practically coincides with its limiting form for $E_i \rightarrow 0$. We also made the Monte-Carlo simulation of 3-Compoton scattering process of classically correlated photon pairs for the idealised experiment with $4\pi$ geometry and  precise resolution of all parameters.  Data points with error bars in  Fig.~\ref{fig:ConcrMC} correspond to twice the derived value of $C$ in this virtual experiment. Now the important point: Eq.~(\ref{eq:CC}) describes the evolution of quantum entanglement within the framework of QFT, while the simulated data in Fig.~\ref{fig:ConcrMC} correspond to classical correlations. They are fundamentally different, e.g., in the measurements of classical correlations there is no place for SAD, which, on the contrary, is considered inherent in quantum mechanics,
giving rise to the EPR paradox.  However, we see that the evolution of both types of correlations during the measurement process is the same, up to a constant factor of two.

\subsection{Polarization state after scattering}

In this section we will obtain a simple and transparent approximation for the polarization state of a photon pair after scattering.  Despite the simplifications, at high energies, $E_i \gtrsim m_e$, the concurrence of the derived state  gives sufficiently accurate approximation of Eq.~(\ref{eq:CC}). 

Let us align two coordinate axis with the scattering plane, the third one will be orthogonal to it (vertical).  Stokes vector of  scattered photon, $\xi_k^f$, is a linear function of the initial one,  $\xi_k^f = \xi_k^f (\xi_k^i )$. Using explicit relations from Ref.~\cite{berestetskii} we find
\begin{equation}
\begin{aligned}
& P^{~}_{V\rightarrow H}=\frac{1-\xi^f_3(+1)}{2}= \frac{\gamma-2}{2\gamma}, \\
& P^{~}_{H\rightarrow V}=\frac{1+\xi^f_3(-1)}{2}= \frac{\gamma-2}{2(\gamma - \sin^2 \theta)}, 
 \end{aligned}
 \label{eq:stokes}
\end{equation}
where $P^{~}_{V\rightarrow H}$ gives the probability to find horizontal polarization of a scattered photon if the  initial one  was vertically polarized,  $P^{~}_{H\rightarrow V}$ is defined similarly, while $P^{~}_{V\rightarrow V} = 1- P^{~}_{V\rightarrow H}$ and $P^{~}_{H\rightarrow H} = 1 - P^{~}_{H\rightarrow V}$.

Following EPR reasoning it was assumed~\cite{Bohm2} that the state of two photons, which were maximally entangled initially, after scattering (measurement) of the first photon will collapse to a mixed separable state described by the density matrix
\begin{equation}
 \rho = \frac{1}{2}( \ket{{H_a}{V_b}}\bra{{V_b}{H_a}} + \ket{{V_a}{H_b}}\bra{{H_b}{V_a}}).
 \label{eq:density_matrix}
\end{equation}
However, since the visibility of any separable state is zero, such an assumption contradicts to the recent measurements  \cite{Abdurashitov,Ivashkin,Parashari2023} and is ruled out. Therefore, we assume the following.  

\begin{figure*}
  \centering
 \includegraphics[width=0.495\textwidth]{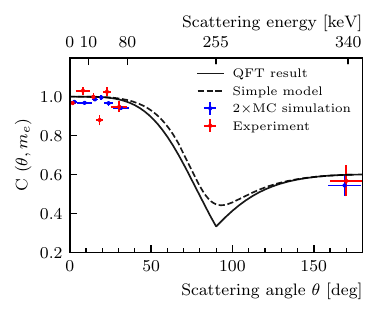}
 \includegraphics[width=0.495\textwidth]{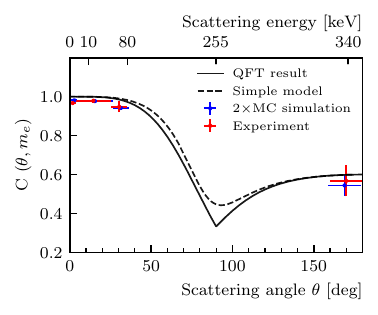}
   \caption{
The solid and dotted curves, as well as the meaning of the blue dots with error bars, are the same as in the left panel of Fig.~\ref{fig:ConcrMC}. On the left,   the red dots with error bars represent the result of our experiment with annihilation photons. On the right, the first four bins with non-zero $\theta$ are combined into one.  The datapoints at $\theta=0$ correspond to the initial annihilation photons that did not experience scattering, and the corresponding symbols are shifted slightly to the right for clarity.
}
\label{fig:ConcrData}
\end{figure*}

A) The scattering (evolution) of the first photon proceeds locally, according to QFT, and does not affect the polarization state of the second photon. There is no steering or ``spooky action'' at a distance here. 

B) The two-photon state remains pure after scattering. This is, of course, an approximation. The state is fixed then by the following ket vector
\begin{equation}
 \begin{aligned}
  {\Psi_2} = & \frac{1}{\sqrt2}( P^{1/2}_{H_a\rightarrow H_a} \ket{{H_a}{V_b}} + P^{1/2}_{V_a\rightarrow V_a} \ket{{V_a}{H_b}} + \\ & +  P^{1/2}_{H_a\rightarrow V_a} \ket{{V_a}{V_b}} + P^{1/2}_{V_a\rightarrow H_a} \ket{{H_a}{H_b}}  ).
\end{aligned}  
\label{eq:wavefunc2}
\end{equation}
Transitions $H \leftrightarrow V$ simply lead to the appearance in the state decomposition of two other basis vectors $\ket{{V_a}{V_b}}$ and $\ket{{H_a}{H_b}}$. The concurrence of resulting state is easy to find~\cite{Wootters2001} and it is equal to
\begin{equation}
C( {\Psi_2}) = \left|{ \sqrt{P^{~}_{H_a\rightarrow H_a} P^{~}_{V_a\rightarrow V_a}} - \sqrt{P^{~}_{H_a\rightarrow V_a}  P^{~}_{V_a\rightarrow H_a}}  }\right|.
 \label{eq:C2}
\end{equation}
This function is shown in Fig.~\ref{fig:ConcrMC} by the dashed line. Note that for forward and backward scattering we have $\sin\theta = 0$, which results in $P^{~}_{H\rightarrow V}  = P^{~}_{V\rightarrow H} $, see Eq.~(\ref{eq:stokes}). This gives $C( {\Psi_2})  = 1 -2P^{~}_{V_a\rightarrow H_a} = 2/\gamma$, which coincides with the forward and backward scattering limits of the exact QFT result Eq.~(\ref{eq:CC}).  Moreover, the simple model is quite accurate for $E_i = m_e$ at all scattering angles, and the agreement with the exact result Eq.~(\ref{eq:CC}) improves at higher energies. This proves that $C(\rho)$ in  Eq.~(\ref{eq:CC}) is the concurrence. The discrepancy at low energies and intermediate angles is due to the pure state approximation, and the decoherence of the off-diagonal terms in $\rho$ should correct this.

\section{Measured Entanglement}

SAD is not expected to arise during the evolution of entanglement within QFT. And this did not happen, as we saw above. If it exists, as is often explicitly or implicitly assumed, then it must be new physics. This should and can be investigated experimentally within our approach.

In the experiment with annihilation photons, we collect both types of events in parallel: the original entangled pairs of photons and those events where one of the photons was previously scattered.  Events are tagged depending on the energy deposited in the intermediate prescatterers and classified by scattering angle, see section~\ref{sec:Selection}. 
The events forming the set of primary annihilation photons did not interact in the prescatterer and passed freely through it. This is the largest group. We divide the events that interacted into bins according to the scattering angle.  The scattering angle is determined by the measured energy release $\Delta E$ in the prescatterer as $\cos\theta = (m_e - 2\Delta E)/ (m_e - \Delta E)$. 

The analysing power of a polarimeter that  measured prescattered photon (let it be photon ``a'') is calculated as $A_a =A(E_{ai}, E_{af})$ via Eqs.~(\ref{eq:sangle}) and (\ref{eq:apow}), where $E_{ai} = m_e - \Delta E$, and  $E_{af}$ is measured by  photon counters. For a given $\theta$ there is a distribution of $A_a$ due to the spread over the second scattering angle $\theta_a$, so we define  $\bar{A_a}$ as the average over all measured $E_{af}$ in the corresponding $\theta$ bin. The analysing power of the other polarimeter is fixed, but its value deviates from  $A(m_e, 90^\circ) = 2/3$ due to the finite size geometry of the scatterers and counters. The corresponding spread of $\theta_b$ around $90^\circ$ produces $\bar{A_b} = 0.661$. We extract visibility for a given bin in $\theta$ by fitting the distributions of  photons detected at counters separated by an angle $\phi$ to  Eq.~(\ref{eq:Sfunction}).  
Further details are given in section~\ref{subsec5-C}. 

Finally, we extract the entanglement measure as $C(\theta) = \nu(\theta) /\bar{A_a}(\theta) /\bar{A_b}$. The results are shown in Fig.~\ref{fig:ConcrData}. We repeated the same procedure for a simulated, initially classically correlated photon data set in the realistic geometry of our experiment using the Geant4 toolkit~\cite{Geant4,Geant2016}. The results of this procedure are also presented in Fig.~\ref{fig:ConcrData}. Again, experimentally measured entanglement as a function of scattering angle is twice the classical correlations within error limits, and both are consistent with our theoretical expectations.

\section{Discussion and outlook}

We investigated the polarization correlations of initially maximally entangled photons in a state where one of them was previously scattered. We did this in three ways: analytically, in Monte-Carlo simulations and in dedicated experiment for annihilation photons. Pre-scattering can be thought of as a measurement of polarization since, e.~g., it is in no way different from the scattering employed in subsequent measurements in Compton polarimeters.

Equation~(\ref{eq:CC})  quantifies entanglement of the photon pair after one of them scatters. Concurrence (\ref{eq:C2}) of the final pure state (\ref{eq:wavefunc2}) at $E_0 \gtrsim m_e$ closely  reproduces  entanglement behaviour as a function of $\theta$ and coincides exactly for forward and backward scattering at all energies. The difference between the pure state approximation and the exact QFT result~(\ref{eq:CC})  at intermediate scattering angles is due to decoherence of the off-diagonal elements of the density matrix describing the initial pure state. We expect that $C(\theta,E_i)$  will still be given by the concurrence for the final mixed state as well.  

In deriving these results for quantum entangled states, we assumed  the absence of steering, or, in other words, the absence of ``spooky action at a distance''. Moreover, we have shown that the initially classically correlated state evolves in exactly the same way, except that the resulting $C(\theta,E_i)$ is always half as large~\footnote{This means that the evolution of entangled states can be modelled using, say, the Geant4 toolkit and initially classically correlated states, and then multiplying the results by a factor of 2.}.  Importantly, $C(\theta,m_e)$ describes our experimental results within errors for measured ranges of scattering angle. Collapse of the two photon state into separable one does not occur despite scattering is a measurement of polarization of one of the photons. All of the above proves the absence of quantum "spooky action at a distance" underlying the EPR paradox. 
 
Having said that, we believe that measuring an intermediate range of scattering angles at other energies using approach described in this paper is still relevant to closing a possible loophole. SAD is sought experimentally. In particular, recent sophisticated experiments have placed limits on the speed of SAD (if it exists) that exceeds the speed of light by at least four orders of magnitude~\cite{Salart2008,SantamariaAmato2023}. Our approach provides a method to experimentally test for the presence of SAD: if a change in the state of a distant photon is needed to explain the measured correlations, it is there, otherwise it is absent.

We also explained and quantified the "decoherence puzzle" observed in~\cite{Abdurashitov,Ivashkin} and confirmed in~\cite{Parashari2023} recently. Namely, the azimuthal angular distributions for maximally entangled and scattered photon pairs are very similar over a wide range of pre-scattering angles up to $\theta=40^\circ$ (see Figs.~\ref{fig:ConcrData}, \ref{fig_entangled} and ~\ref{fig_R_decoh_multi}). This effect is of particular importance in the development of quantum entangled PET~\cite{watts,McNamara,Toghyani,kozuljevich}, based on the hypothetically assumed absence of correlations for pre-scattered events, and means that the corresponding kinematic constraints cannot be used to suppress scattering background in QE-PET.

\section{Methods}
\label{sec:methods}

\subsection{Experiment}
\label{sec:Experiment}

\begin{figure*}[htpb]%
  \centering
 \includegraphics[width=0.49\textwidth]{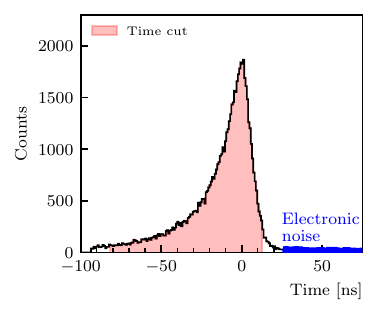}
 \includegraphics[width=0.5\textwidth]{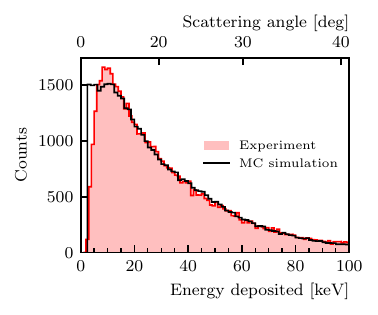}
   \caption{
Left:  Time spectrum of events in the intermediate scatterer.  The flat electronic noise spectrum is indicated by the blue line. The time window selected for analysis is displayed as a shaded area. Right: The energy release spectrum in the intermediate scatterer  is shown by the  red line. The black line corresponds to the simulated Monte Carlo energy spectrum. 
}\label{fig_time_spectra}
\end{figure*}

To measure the annihilation photons in maximally entangled (initial) and pre-scattered states the dedicated  experimental setup \cite{Abdurashitov} was constructed. The principles of its operation are illustrated in Fig.~\ref{scheme_polarimeter}. Two identical arms of Compton polarimeters are mounted on opposite sides of a $^{22}$Na positron source embedded in a 1 mm thick aluminum plate.
It is known~\cite{Garwin}, that positrons in aluminium practically immediately thermalize and annihilate with electrons with zero relative angular momentum. Therefore, the produced pairs of annihilation photons are in maximally entangled state~\cite{Bohm,Snyder}.

The source is located in the center of a lead cube having a horizontal tunnel with a diameter of 5 mm. This provides collimated beams of annihilation photons in opposite directions from the source, aimed at the main scatterers of the Compton polarimeters.  A photon pair is registered if both main scatterers detect signals from recoil electrons within specified small time window. We placed pre-scatterers  in front of the main scatterers.  Polarimeters measure the polarization of either the initial entangled photons, if no interaction is detected in intermediate scatterers, or  the photons pre-scattered on ambient electrons in one of the intermediate scatterers, otherwise.  Both types of events were collected in parallel and classified by scattering angle depending on the energy release in the counters  and intermediate scatterers.  Simultaneous signal events in both pre-scatterers are rare and we do not take them into account at this stage. 

Scattered photons are recorded by a ring of 16 scintillation counters installed in each arm orthogonally to the setup axis with an azimuthal angle between adjacent counters of $22.5^{\circ}$.  This azimuthally symmetrical arrangement cancels possible systematic errors caused by different efficiencies and inaccuracies in the position of photon counters. Also, a complete ring of scattered photon detectors makes it possible to measure scattered photon correlations over the entire range of azimuthal angles. Data were collected over a six-month experiment.

Along with the real one, a virtual experiment was conducted on a simulated model of the experimental setup. The Monte Carlo simulation of experiment is based on the latest version of Geant4~\cite{Geant4, Geant2016}  particle simulation toolkit (with physics list “G4EmLivermorePolarized”), commonly used in nuclear and high energy physics. 

\subsection{Selection and classification of events}
\label{sec:Selection}

The initial and pre-scattered states of the annihilation photons are separated from each other depending on the detection of pre-scattering energy release, if any.  Therefore, reliable identification of photon interactions in an intermediate scatterer is mandatory for tagging quantum states. In our case a bright inorganic scintillator, a gadolinium-aluminum-gallium garnet (GAGG), is used as an intermediate scatterer, providing a low detection threshold of deposited energies. 

The amplitude and timing of the signals make it possible to reliably reconstruct the interaction in this scintillator. Corresponding spectra are presented in Fig.~\ref{fig_time_spectra}. Since time reconstruction depends on energy deposition, the time spectrum (Fig.~\ref{fig_time_spectra}, left) has an asymmetric shape with a tail corresponding to low energy deposition. Accordingly, a narrow time window would discard a significant portion of low-energy events, resulting in a distorted energy spectrum compared to the simulated data. To avoid these shortcomings, a wide time window was used, indicated in Fig.~\ref{fig_time_spectra} as the shaded area. As can be seen from the shape of the simulated energy spectrum, only a very small fraction of pre-scattered events with the lowest energy release in the intermediate scatterer are lost.
 
\begin{figure}[htpb]%
  \centering
  \includegraphics[width=0.5\textwidth]{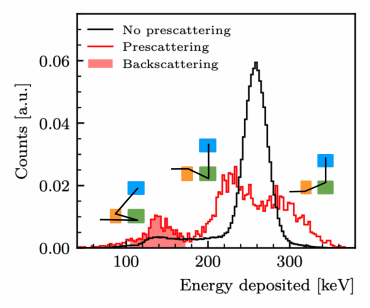}
  \caption{
Energy deposition spectrum in  photon counters for three types of events. The black line corresponds to correlated events with zero energy release in both intermediate scatterers. The red line corresponds to events in which one of the photons experienced preliminary scattering. The different scattering kinematics are shown near the corresponding peaks, where the orange, green, and blue boxes represent the intermediate scatterer, the main scatterer, and the  photon counter, respectively. The energy range corresponding to backscattering events is highlighted in red shading.
}\label{fig_backscattering}
\end{figure}

On the other hand, some fraction of the initial photons can be accepted as pre-scattered events if the relatively high electronic noise (with equivalent energy of about a few keV) in the intermediate scatterer mimics the interaction signal. In this case, the electronic noise timing will have a flat distribution, see time spectrum. The fraction of such noisy events in a wide time window is at the sub-percentage level, nevertheless we cut these events adjusting the time window as shown.

The measured energy deposit $\Delta E$ allows us to determine the final photon energy after the first scattering as $E_f = m_e - \Delta E$ and determine the scattering angle using Eq.~(\ref{eq:sangle}). The resulting scattering angle is displayed on the top horizontal axis in Fig.~\ref{fig_time_spectra}, right panel, and is used as an argument of $C(\theta, m_e)$ in Fig.~\ref{fig:ConcrData}. 
The accuracy of the scattering angle determination is related to the energy resolution of the GAGG scintillator and is equal to
$
{\Delta \theta }/{\theta} = {0.05}/{\sqrt{E_G/(30\; \rm keV)}} ,
$
where $E_G$ is the energy released in  GAGG.
 
Energy deposits are further measured in the polarimeter - in the main scatterer and in NaI photon counters.  Note that $E_f$ after the pre-scatterer is equal to $E_{ia}$ for the polarimeter.  This allows us to determine the analysing powers $A(E_{ai},E_{af})$ and  $A(E_{bi},E_{bf})$ according to the Eqs.~(\ref{eq:apow}) and (\ref{eq:sangle}), which enter Eq.~(\ref{eq:MainRel}).

The energy spectra measured in photon counters are presented in Fig. \ref{fig_backscattering} for several event classes. Since in Compton scattering at 90$^\circ$ the recoil electron receives half the initial energy of the photon, the scattered photon will also have an energy of about 255 keV, as seen in the black spectrum in Fig. \ref{fig_backscattering} which corresponds to the direct annihilation photon. The low-energy tail in this spectrum corresponds to the partial energy release in the counter caused by the photon escape from it. 

The red line in Fig. \ref{fig_backscattering} shows the energy spectrum of events with pre-scattering energy release in the range of [30,60] keV.  The structure of  peaks reflects pre-scattering kinematics with different scattering angles. The two rightmost peaks correspond to the second scattering located in the prescattering plane with $\theta_a > 90^\circ$ (peak at $E_{af} \approx 220$ keV)  and $\theta_a < 90^\circ$ (peak at $E_{af} \approx 280$ keV), the minimum between them corresponds to the second scattering in the orthogonal direction, $\theta_a = 90^\circ$.  
The shaded region of the red spectrum below the lowest energy peak corresponds to events with a pre-scattering angle of about 180$^\circ$. These backscattered events undergo the prior scattering in the main scatterer, while the polarization measurement is done in the intermediate scatterer. 

\subsection{Measured visibility}
\label{subsec5-C}

\begin{figure}[htpb]%
  \centering
  \includegraphics[width=.48\textwidth]{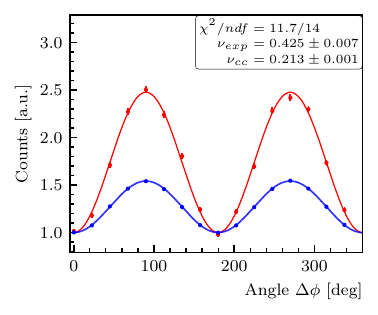}
  \includegraphics[width=.48\textwidth]{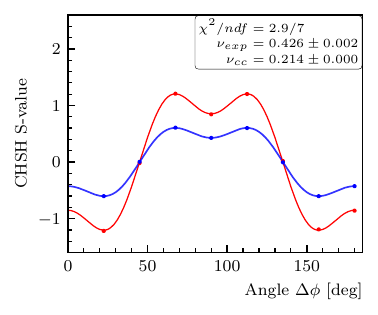}
\caption{
Measured azimuthal correlations of initial annihilation photons. In both figures, the red lines show the fit to the experimental data (which are represented by points with error bars).  Blue lines and symbols correspond to Monte Carlo simulations of initially classically correlated photons. Top: the fit function Eq.~(\ref{eq:vis_def}) is used.  Bottom: fitting by the  $S$-function Eq.~(\ref{eq:Sfunction}).}
\label{fig_entangled}
\end{figure}

Visibility calculation is straightforward. Simultaneous activation of counters (within a given time window) in each polarimeter arm is recorded as an event that is added to the bin corresponding to the azimuthal angle  $\Delta \phi$ between the counters. The resulting distribution is then fitted by Eq.~(\ref{eq:vis_def}) if  simple procedure is used. Events in pairs with azimuthal  separations that are multiples of $22.5^\circ$ in 16 elementary Compton polarimeters on each side of the setup are summed for all relevant combinations of polarimeters in the fit by the CHSH S-function Eq.~(\ref{eq:Sfunction}), for details see~\cite{Abdurashitov}. The result for direct annihilation photons is shown in Fig.~\ref{fig_entangled}, both for the real experiment and for the MC simulation of initially classically correlated photons.  The accuracy of determining $\nu$ by the CHSH method is approximately 3 times higher, which is confirmed by the corresponding values of $\chi^2/\rm ndf $.  Therefore, we use this method in calculation of $C(\theta)$.  Note that the visibility for classically correlated photons (i.e., with random but mutually orthogonal initial polarization directions) is  2 times lower compared to the maximally entangled quantum state, in agreement with the analytical expectation for direct photons.

The  same procedures were repeated for the case when one of the photons experienced prior scattering. The energy $\Delta E$ released in the first scattering is determined by the scattering angle. In our setup we can assess $\theta$ up to $35^\circ$ for forward scattering. We divided this range into 5 bins with a comparable number of events in each. On the other hand, all backscattered events  were combined into single bin due to statistical limitations here. This is quite legitimate since $C(\theta)$ is expected to be constant over the entire available range $160^\circ < \theta < 180^\circ$ anyway, see Fig.~\ref{fig:ConcrMC}. Corresponding angular correlation functions  are shown in Figs.~\ref{fig_R_decoh_multi} and \ref{fig_S_decoh_multi}.

\begin{widetext}

\begin{figure}[hbtp]%
\centering
\includegraphics[width=0.85\textwidth]{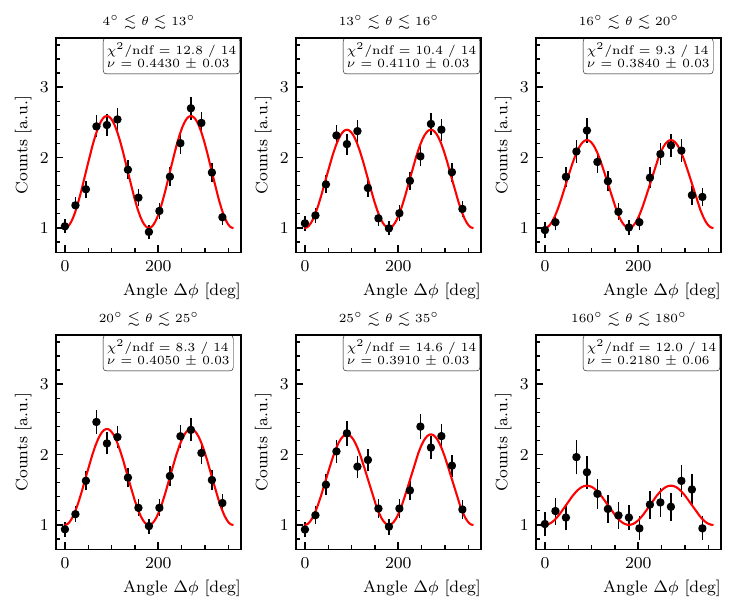}
\caption{
Normalized counts in detectors of scattered photons as a function of azimuthal angle between these photons for different energies of recoil electrons. The red solid lines correspond to the fitting function Eq.~(\ref{eq:vis_def}).  Derived values of visibility are shown in each histogram. 
}
\label{fig_R_decoh_multi}
\end{figure}

\begin{figure}[hbtp]%
\centering
\includegraphics[width=0.85\textwidth]{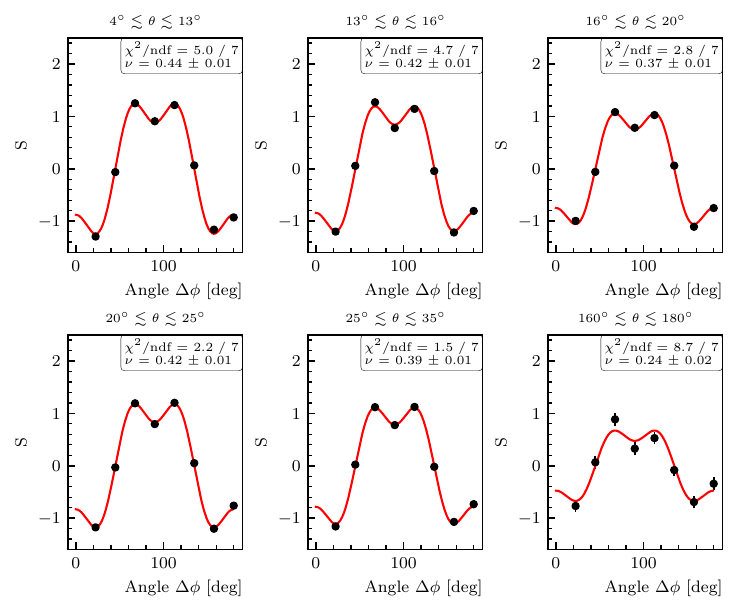}
\caption{
Dependence of the correlation $S$-function on the azimuthal angle between polarimeters for different scattering energies. The red solid lines correspond to the fitting function Eq.~\ref{eq:Sfunction}. Derived values of visibility are shown in each histogram.
}
\label{fig_S_decoh_multi}
\end{figure}

\end{widetext}

\subsection{Factorisation of visibility}
\label{sec:Fact}

According to Ref.~\cite{Caradonna2024}, the 3-Compton scattering probability for a maximally entangled photon pair, up to angles independent factors, is given by
\begin{equation}
    \begin{aligned}
       & P(\phi_{b},\phi_{a}) = A\left(\theta_{b},\theta_{a};\theta \right) -  \\ & - B\left(\theta_{b},\theta_{a}\right)C\left(\phi_{b},\phi_{a};\theta\right) + D\left(\theta_{b},\theta_{a},\phi_{b},\phi_{a};\theta \right),
    \end{aligned}
\label{eqn:31}
\end{equation}
where we used our notations for the scattering angles, namely $\theta_b$, $\theta_a$, $\theta$, and $\phi_{b}$, $\phi_{a}$ are azimuthal angles of the scattered photons $b$ and $a$ in a properly defined $\theta$-dependent global coordinate system, see~\cite{Caradonna2024}, and in $P$ we indicate only azimuthal angles as an arguments. Explicit expressions for $A\left(\theta_{b},\theta_{a};\theta \right)$ and  $B\left(\theta_{b},\theta_{a};\theta\right)$ don't matter at the moment, while
    \begin{equation}
        \begin{aligned}
	   C\left(\phi_{b},\phi_{a};\theta\right) = \cos2\phi_{b}\cos2\phi_{a} - \cos\theta \sin2\phi_{b}\sin2\phi_{a},
        \end{aligned}
    \label{eqn:32c}	
    \end{equation}
and
    \begin{equation}
        \begin{aligned}
		& D\left(\theta_{b},\theta_{a},\phi_{b},\phi_{a};\theta \right)  = \\ & = \frac{1}{2} B\left(\theta_{b},\theta_{a}\right)  \sin^2\theta  \cos2\phi_{b} \cos2\phi_{a}  + (\dots)
	\end{aligned}
    \label{eqn:32d}	
    \end{equation}
where $(\dots)$ denote terms linear in $\cos2\phi_{b}$ or $\cos2\phi_{a}$ which integrate to zero in what follows.  
Now comes the key point.  We do not have a preferred direction, so the argument of the correlation function R must be the azimuthal angle $\phi$ between final scattering planes of photons $b$ and $a$,  regardless of the specific values of $\phi_a$ or $\phi_b$. Namely, for forward scattering $\phi =  \phi_a + \phi_b$, see Fig.~\ref{scheme_polarimeter}, which results in
\begin{equation}
R(\phi) = \frac{1}{\pi}\int_0^{\pi} d\phi_b P(\phi-\phi_b,\phi_b) .
\end{equation}
For backscattering $\phi = \phi_b - \phi_a$ to preserve the orientation of the basis and
\begin{equation}
R(\phi) = \frac{1}{\pi}\int_0^{\pi} d\phi_b P(\phi_b-\phi,\phi_b) .
\end{equation}
Integration gives
\begin{equation}
 \begin{aligned}
&R(\phi) =  A\left(\theta_{b},\theta_{a};\theta \right) - \\ & - B\left(\theta_{b},\theta_{a}\right)\left[1+ |\cos\theta | - \frac{1}{2}   \sin^2\theta \right] \frac{\cos2\phi}{2} \, .
 \end{aligned}
\end{equation}
Therefore
\begin{equation}
\nu = \frac{B\left(\theta_{b},\theta_{a}\right)}{2A\left(\theta_{b},\theta_{a};\theta \right)}  \left[1+ |\cos\theta | - \frac{1}{2}   \sin^2\theta \right]  .
\label{eq:nufin}
\end{equation}
Now, in notations of Ref.~\cite{Caradonna2024} and using our scattering angle notations
    \begin{equation}
        B\left(\theta_{b},\theta_{a}\right) = 2t^{(11)}_{12}t^{(22)}_{12} = 2\sin^2\theta_b \,\sin^2\theta_a,
    \label{eqn:32b}
    \end{equation}
 \begin{equation}
        A\left(\theta_{b},\theta_{a};\theta_{}\right) = t^{(11)}_{11}t^{(21)}_{11}t^{(22)}_{11} .
    \label{eqn:32a}
    \end{equation}
 Each  matrix element $ t^{(pj)}_{11}$ has the following structure  
$$
     t^{(pj)}_{11} = 1+\cos^{2}\theta_{pj}+\left(E_{p(j-1)}-E_{pj}\right)\left(1-\cos\theta_{pj}\right), 	
$$
where $p$ refers to particle $a=2$ or $b=1$, and $j$ denotes its first or second scattering. Note that $\Delta E_{pj} = E_{p(j-1)}-E_{pj}$ is the energy release in scattering $j$.  Noticing that $\Delta E (1-\cos\theta ) = \gamma - 2$ we find
$$
\begin{aligned}
& t^{(11)}_{11} = \gamma_b^{~} - \sin^2\theta_b,  \qquad  t^{(22)}_{11} = \gamma_a - \sin^2\theta_a, \\ &  t^{(21)}_{11} = \gamma - \sin^2\theta \, ,
\end{aligned}
$$
where each $\gamma$, Eq.~(\ref{eq:sangle}), uses the initial and final energies corresponding to a specific scattering. We see that the visibility Eq.~(\ref{eq:nufin}) for arbitrary $E_i$ is decomposed into the analyzing power of each polarimeter, Eq.~(\ref{eq:apow}), and  the entanglement  measure of the state arising after pre-scattering, which is given by Eq.~(\ref{eq:CC}).

\subsection{Classically correlated photons}
\label{sec:CC}

Assume we organise a source which emits randomly polarized photons in the direction $+z$ with uniform probability distribution. For each such photon with an azimuthal polarization angle $\phi$, the source emits a photon in the $-z$ direction and with a polarization $\phi + 90^\circ$. Each of the photons scatters in respective Compton polarimeter.
According to Eq.~(\ref{eq:kl-ni}) probability to detect photon $a$ at the direction $\theta_a$ forming scattering plane at $\phi_a$ with respect to the original polarization direction is
\begin{equation}
P_a = \gamma_a - 2\sin^2{\theta_a} \cos^2{\phi_a}. \nonumber
\end{equation}
Similarly for the photon $b$
\begin{equation}
P_b = \gamma_b - 2\sin^2{\theta_b} \cos^2(\phi_b + 90^\circ). \nonumber
\end{equation}
Both scattering planes will be parallel if $\phi_b=\phi_a$ or $\phi_b=\phi_a+180^\circ$. We find
\begin{equation}
P_{\parallel} = \frac{1}{\pi} \int_0^{2\pi} d\phi\ ( \gamma_a - 2\sin^2{\theta_a} \cos^2 \phi ) ( \gamma_b - 2\sin^2{\theta_b} \sin^2 \phi ). \nonumber
\end{equation}
Similarly, scattering planes will be orthogonal if $\phi_b=\phi_a \pm 90^\circ$ giving
\begin{equation}
P_{\perp} = \frac{1}{\pi} \int_0^{2\pi} d\phi\ ( \gamma_a - 2\sin^2{\theta_a} \cos^2 \phi ) ( \gamma_b - 2\sin^2{\theta_b} \cos^2 \phi ). \nonumber
\end{equation}
After integrations we find
\begin{equation}
 \begin{aligned}
& P_{\perp} - P_{\parallel} = 2 \sin^2{\theta_a} \sin^2{\theta_b},  \\ & P_{\perp} + P_{\parallel} = 4 (\gamma_a -  \sin^2{\theta_a}) (\gamma_b - \sin^2{\theta_b}) , \nonumber
\end{aligned}
\end{equation}
which gives
\begin{equation}
\nu =  \frac{P_{\perp} - P_{\parallel}}{P_{\perp} + P_{\parallel}}  =   \frac{1}{2} \, A(E_{ai},\theta_{a})\, A(E_{bi},\theta_{b}) .
  \nonumber
\end{equation}

\section{Acknowledgements}

The authors are grateful to D. Levkov, M. Libanov, and A. Panin for useful discussions.

\bibliography{sample}

\end{document}